\input amssym.def
\input epsf


\magnification=\magstephalf
\hsize=14.0 true cm
\vsize=19 true cm
\hoffset=1.0 true cm
\voffset=2.0 true cm

\abovedisplayskip=12pt plus 3pt minus 3pt
\belowdisplayskip=12pt plus 3pt minus 3pt
\parindent=1.0em


\font\sixrm=cmr6
\font\eightrm=cmr8
\font\ninerm=cmr9

\font\sixi=cmmi6
\font\eighti=cmmi8
\font\ninei=cmmi9

\font\sixsy=cmsy6
\font\eightsy=cmsy8
\font\ninesy=cmsy9

\font\sixbf=cmbx6
\font\eightbf=cmbx8
\font\ninebf=cmbx9

\font\eightit=cmti8
\font\nineit=cmti9

\font\eightsl=cmsl8
\font\ninesl=cmsl9

\font\sixss=cmss8 at 8 true pt
\font\sevenss=cmss9 at 9 true pt
\font\eightss=cmss8
\font\niness=cmss9
\font\tenss=cmss10

 at 12 true pt
\font\bigrm=cmr10 at 12 true pt
 at 12 true pt

 at 14 true pt
\font\Bigrm=cmr12 at 16 true pt
 at 14 true pt

\font\SBbb=msbm9

\catcode`@=11
\newfam\ssfam

\def\tenpoint{\def\rm{\fam0\tenrm}%
    \textfont0=\tenrm \scriptfont0=\sevenrm \scriptscriptfont0=\fiverm
    \textfont1=\teni  \scriptfont1=\seveni  \scriptscriptfont1=\fivei
    \textfont2=\tensy \scriptfont2=\sevensy \scriptscriptfont2=\fivesy
    \textfont3=\tenex \scriptfont3=\tenex   \scriptscriptfont3=\tenex
    \textfont\itfam=\tenit                  \def\it{\fam\itfam\tenit}%
    \textfont\slfam=\tensl                  \def\sl{\fam\slfam\tensl}%
    \textfont\bffam=\tenbf \scriptfont\bffam=\sevenbf
    \scriptscriptfont\bffam=\fivebf
                                            \def\bf{\fam\bffam\tenbf}%
    \textfont\ssfam=\tenss \scriptfont\ssfam=\sevenss
    \scriptscriptfont\ssfam=\sevenss
                                            \def\ss{\fam\ssfam\tenss}%
    \normalbaselineskip=13pt
    \setbox\strutbox=\hbox{\vrule height8.5pt depth3.5pt width0pt}%
    \let\big=\tenbig
    \normalbaselines\rm}

\def\ninepoint{\def\rm{\fam0\ninerm}%
    \textfont0=\ninerm      \scriptfont0=\sixrm
                            \scriptscriptfont0=\fiverm
    \textfont1=\ninei       \scriptfont1=\sixi
                            \scriptscriptfont1=\fivei
    \textfont2=\ninesy      \scriptfont2=\sixsy
                            \scriptscriptfont2=\fivesy
    \textfont3=\tenex       \scriptfont3=\tenex
                            \scriptscriptfont3=\tenex
    \textfont\itfam=\nineit \def\it{\fam\itfam\nineit}%
    \textfont\slfam=\ninesl \def\sl{\fam\slfam\ninesl}%
    \textfont\bffam=\ninebf \scriptfont\bffam=\sixbf
                            \scriptscriptfont\bffam=\fivebf
                            \def\bf{\fam\bffam\ninebf}%
    \textfont\ssfam=\niness \scriptfont\ssfam=\sixss
                            \scriptscriptfont\ssfam=\sixss
                            \def\ss{\fam\ssfam\niness}%
    \normalbaselineskip=12pt
    \setbox\strutbox=\hbox{\vrule height8.0pt depth3.0pt width0pt}%
    \let\big=\ninebig
    \normalbaselines\rm}

\def\eightpoint{\def\rm{\fam0\eightrm}%
    \textfont0=\eightrm      \scriptfont0=\sixrm
                             \scriptscriptfont0=\fiverm
    \textfont1=\eighti       \scriptfont1=\sixi
                             \scriptscriptfont1=\fivei
    \textfont2=\eightsy      \scriptfont2=\sixsy
                             \scriptscriptfont2=\fivesy
    \textfont3=\tenex        \scriptfont3=\tenex
                             \scriptscriptfont3=\tenex
    \textfont\itfam=\eightit \def\it{\fam\itfam\eightit}%
    \textfont\slfam=\eightsl \def\sl{\fam\slfam\eightsl}%
    \textfont\bffam=\eightbf \scriptfont\bffam=\sixbf
                             \scriptscriptfont\bffam=\fivebf
                             \def\bf{\fam\bffam\eightbf}%
    \textfont\ssfam=\eightss \scriptfont\ssfam=\sixss
                             \scriptscriptfont\ssfam=\sixss
                             \def\ss{\fam\ssfam\eightss}%
    \normalbaselineskip=10pt
    \setbox\strutbox=\hbox{\vrule height7.0pt depth2.0pt width0pt}%
    \let\big=\eightbig
    \normalbaselines\rm}

\def\tenbig#1{{\hbox{$\left#1\vbox to8.5pt{}\right.\n@space$}}}
\def\ninebig#1{{\hbox{$\textfont0=\tenrm\textfont2=\tensy
                       \left#1\vbox to7.25pt{}\right.\n@space$}}}
\def\eightbig#1{{\hbox{$\textfont0=\ninerm\textfont2=\ninesy
                       \left#1\vbox to6.5pt{}\right.\n@space$}}}

\font\sectionfont=cmbx10
\font\subsectionfont=cmti10

\def\figurecaptionfont{\ninepoint}
\def\tablecaptionfont{\ninepoint}
\def\footnotefont{\eightpoint}


\newcount\equationno
\newcount\bibitemno
\newcount\figureno
\newcount\tableno

\equationno=0
\bibitemno=0
\figureno=0
\tableno=0


\footline={\ifnum\pageno=0{\hfil}\else
{\hss\rm\the\pageno\hss}\fi}


\def\section #1. #2 \par
{\vskip0pt plus .20\vsize\penalty-100 \vskip0pt plus-.20\vsize
\vskip 1.6 true cm plus 0.2 true cm minus 0.2 true cm
\global\def\equationlabel{#1}
\global\equationno=0
\leftline{\sectionfont #1. #2}\par
\immediate\write\terminal{Section #1. #2}
\vskip 0.7 true cm plus 0.1 true cm minus 0.1 true cm
\noindent}


\def\subsection #1 \par
{\vskip0pt plus 0.8 true cm\penalty-50 \vskip0pt plus-0.8 true cm
\vskip2.5ex plus 0.1ex minus 0.1ex
\leftline{\subsectionfont #1}\par
\immediate\write\terminal{Subsection #1}
\vskip1.0ex plus 0.1ex minus 0.1ex
\noindent}


\def\appendix #1 \par
{\vskip0pt plus .20\vsize\penalty-100 \vskip0pt plus-.20\vsize
\vskip 1.6 true cm plus 0.2 true cm minus 0.2 true cm
\global\def\equationlabel{\hbox{\rm#1}}
\global\equationno=0
\leftline{\sectionfont Appendix #1}\par
\immediate\write\terminal{Appendix #1}
\vskip 0.7 true cm plus 0.1 true cm minus 0.1 true cm
\noindent}



\def\equation#1{$$\displaylines{\qquad #1}$$}
\def\enum{\global\advance\equationno by 1
\hfill\llap{(\equationlabel.\the\equationno)}}
\def\noenum{\hfill}
\def\next#1{\cr\noalign{\vskip#1}\qquad}


\def\ifundefined#1{\expandafter\ifx\csname#1\endcsname\relax}

\def\ref#1{\ifundefined{#1}?\immediate\write\terminal{unknown reference
on page \the\pageno}\else\csname#1\endcsname\fi}

\newwrite\terminal
\newwrite\bibitemlist

\def\bibitem#1#2\par{\global\advance\bibitemno by 1
\immediate\write\bibitemlist{\string\def
\expandafter\string\csname#1\endcsname
{\the\bibitemno}}
\item{[\the\bibitemno]}#2\par}

\def\beginbibliography{
\vskip0pt plus .15\vsize\penalty-100 \vskip0pt plus-.15\vsize
\vskip 1.2 true cm plus 0.2 true cm minus 0.2 true cm
\leftline{\sectionfont References}\par
\immediate\write\terminal{References}
\immediate\openout\bibitemlist=biblist
\frenchspacing\parindent=1.8em
\vskip 0.5 true cm plus 0.1 true cm minus 0.1 true cm}

\def\endbibliography{
\immediate\closeout\bibitemlist
\nonfrenchspacing\parindent=1.0em}


\def\figurecaption#1{\global\advance\figureno by 1
\narrower\figurecaptionfont
Fig.~\the\figureno. #1}

\def\tablecaption#1{\global\advance\tableno by 1
\vbox to 0.5 true cm { }
\centerline{\tablecaptionfont%
Table~\the\tableno. #1}
\vskip-0.4 true cm}

\def\thicktablerule{\hrule height1pt}
\def\thintablerule{\hrule height0.4pt}

\tenpoint

\immediate\openin\bibitemlist=biblist
\ifeof\bibitemlist\immediate\closein\bibitemlist
\else\immediate\closein\bibitemlist
\input biblist \fi


\def\thismonth{\ifcase\month\or
January\or February\or March\or April\or May\or June\or
July\or August\or September\or October\or November\or December\fi}



\def\rmd{{\rm d}}
\def\rmD{{\rm D}}
\def\rme{{\rm e}}


\def\gz{{\Bbb Z}}


\def\proof{\noindent{\sl Proof:}\kern0.6em}

\def\frac#1#2{\hbox{$#1\over#2$}}
\def\dual{\mathstrut^*\kern-0.1em}

\def\lvec#1{\setbox0=\hbox{$#1$}
    \setbox1=\hbox{$\scriptstyle\leftarrow$}
    #1\kern-\wd0\smash{
    \raise\ht0\hbox{$\raise1pt\hbox{$\scriptstyle\leftarrow$}$}}
    \kern-\wd1\kern\wd0}
\def\rvec#1{\setbox0=\hbox{$#1$}
    \setbox1=\hbox{$\scriptstyle\rightarrow$}
    #1\kern-\wd0\smash{
    \raise\ht0\hbox{$\raise1pt\hbox{$\scriptstyle\rightarrow$}$}}
    \kern-\wd1\kern\wd0}


\def\nabstar#1{{\nabla\kern0.5pt\smash{\raise 4.5pt\hbox{$\ast$}}
               \kern-5.5pt_{#1}}}

\def\drvstar#1{{\partial\kern0.5pt\smash{\raise 4.5pt\hbox{$\ast$}}
               \kern-6.0pt_{#1}}}

\def\ldrvstar#1{{\lvec{\,\partial}\kern-0.5pt\smash{\raise 4.5pt\hbox{$\ast$}}
               \kern-5.0pt_{#1}}}


\def\GeV{{\rm GeV}}

\def\fm{{\rm fm}}




\def\diracstar#1#2{
    \setbox0=\hbox{$\gamma$}\setbox1=\hbox{$\gamma_{#1}$}
    \gamma_{#1}\kern-\wd1\kern\wd0
    \smash{\raise4.5pt\hbox{$\scriptstyle#2$}}}


\def\tr{{\rm tr}}

\def\Ad{{\rm Ad}\kern0.1em}


\def\twolink{{\Bbb T}}
\def\lineop{{\Bbb L}}
\def\stwolink{\hbox{\SBbb T}}
\def\cc{\smash{\raise5.5pt\hbox{$\scriptstyle\ast$}}\kern-4pt}
\def\sub{{\rm sub}}
\rightline{CERN-TH/2001-208}
\rightline{UT CCP-P-110}
\rightline{MPI-PhT/2001-25}

\vskip 1.3 true cm 
\centerline
{\Bigrm  Locality and exponential error reduction in}
\vskip 1.6ex
\centerline
{\Bigrm  numerical lattice gauge theory}
\vskip 0.6 true cm
\centerline{\bigrm Martin L\"uscher\kern1pt%
\footnote{${\vrule height7.0pt depth1.5pt width0pt}^{\rm a}$}
{\footnotefont%
On leave from Deutsches Elektronen-Synchrotron DESY, 
D-22603 Hamburg, Germany}
}
\vskip1ex
\centerline{\it CERN, Theory Division} 
\centerline{\it CH-1211 Geneva 23, Switzerland}
\vskip 0.4 true cm
\centerline{\bigrm Peter Weisz\kern1pt%
\footnote{${\vrule height6.0pt depth1.5pt width0pt}^{\rm b}$}
{\footnotefont%
Permanent address: Max-Planck-Institut f\"ur Physik,
D-80805 M\"unchen, Germany}
}
\vskip1ex
\centerline{\it Center for Computational Physics, University of Tsukuba}
\centerline{\it Tsukuba, Ibaraki 305-8577, Japan}
\vskip 0.8 true cm
\thintablerule
\vskip 2.0ex
\ninepoint
\leftline{\bf Abstract}
\vskip 1.0ex\noindent
In non-abelian gauge theories without matter fields,
expectation values of large Wilson loops 
and loop correlation functions are difficult to compute 
through numerical simulation, because the signal-to-noise ratio
is very rapidly decaying for increasing loop sizes.
Using a multilevel scheme that exploits the
locality of the theory, we show that the statistical errors
in such calculations can be exponentially reduced.
We explicitly demonstrate this in the SU(3) theory, 
for the case of the Polyakov loop correlation function,
where the efficiency of the simulation
is improved by many orders of magnitude 
when the area bounded by the loops exceeds $1\,\fm^2$.

\vskip 2.0ex
\thintablerule
\vskip -3.0ex 
\tenpoint

\section 1. Introduction

The euclidean expectation values of Wilson loops and their products
are probably the most natural quantities to consider 
in non-abelian gauge theories.
They encode much of the physical information in these theories,
such as the particle spectrum, for example,
and the strength of the force between static colour sources.
These pro\-perties can, however, only be extracted reliably if
one is able to calculate the relevant expectation values accurately
over a significant range of loop sizes and distances.

In lattice gauge theory the computation of
loop expectation values through numerical simulation is in principle
straightforward. The problem is that the signal-to-noise
ratio is exponentially decreasing for large loops,
roughly proportionally to $\rme^{-\sigma A}$ in the confinement phase,
where $\sigma$ denotes the string tension
and $A$ the minimal area spanned by the loop.
According to this law (and if we set $\sigma\simeq1\,\GeV/\fm$),
an increase in $A$ by $1\,\fm^2$ at fixed relative errors requires
the statistics to be multiplied by about $3\times10^{4}$.
Computers are rapidly becoming faster,
but it is clear that a brute-force approach will not pay
under these conditions, i.e.~progress in this field has to come 
mainly from better algorithms and computational strategies.

A significant improvement is achieved, for example, by 
the ``multihit'' method [\ref{MultiHit}], 
where the gauge field variables
in the Wilson loop are replaced by their 
average in the background of all other field variables.
This has no effect on 
the loop expectation value,
but the statistical errors are reduced by an exponential factor with 
exponent proportional to the length of the loop.
Link-blocking techniques [\ref{LBTeper},\ref{LBAlbanese}] and
variational methods [\ref{VMMichael},\ref{VMLuscher}] are further
improvements that are known to be effective in practice and are
widely used. 

In spite of all these advances, it remains difficult to
reach loop areas $A$ greater than about $1\,\fm^2$. 
Moreover, as is generally
the case when applying variational tech\-niques, 
the calculation is biased to some extent by the
choice of basis operators. In the presence of matter fields, for
example, string breaking 
is not observed unless basis elements for both the string and the two-meson
states are included [\ref{SBWittig}--\ref{SBForcrand}].

In this paper we describe a multilevel simulation scheme that leads
to an exponential error reduction with exponent approximately
proportional to the area $A$. 
The algorithmic idea is essentially the same as
in the multihit method [\ref{MultiHit}] but is applied
to pairs of links instead of single links.
It is then possible to iterate the procedure
according to a hierarchical scheme that builds up averages over
increasingly large sublattices. This is not as complicated as it sounds,
and if use is made of recursive functions (programs that call themselves), 
the algorithm is in fact easy to program.

\vskip-3.5ex 

\section 2. Preliminaries

For clarity we shall only consider the case of 
the pure SU(3) gauge theory in this paper, even though
the techniques discussed later are expected to 
be more widely applicable.
The theory is set up on a 4-dimensional hypercubic lattice
with spacing $a$, time-like extent $T$ and spatial size $L$
in the usual way. In particular,
the gauge field is represented by link variables
$U(x,\mu)$ with values in SU(3).
We impose periodic boundary
conditions in all directions and take the standard expression
[\ref{Wilson}]
\equation{
  S[U]={1\over g_0^2}\sum_{x,\mu,\nu}
  \tr\left\{1-
  U(x,\mu)U(x+a\hat{\mu},\nu)U(x+a\hat{\nu},\mu)^{-1}U(x,\nu)^{-1}
  \right\}
  \enum
}
for the lattice action,
where $g_0$ denotes the bare coupling and 
$\hat{\mu}$ the unit vector in direction $\mu$.

For any oriented closed curve $\cal C$ on the lattice,
the associated Wilson loop 
\equation{
  W({\cal C})=\tr\{U({\cal C})\}
  \enum
}
is the trace of the ordered product $U({\cal C})$ of the link variables along
the curve. In the special case of a
straight line that passes through the point $x$ 
and winds once around the lattice in the negative time direction,
$W({\cal C})$ is referred to as a Polyakov loop 
and is denoted by $P(x)$. 

The expectation value of any product ${\cal O}$ of Wilson loops
is defined by
\equation{
  \langle{\cal O}\rangle=
  {1\over{\cal Z}}\int\rmD[U]\,{\cal O}\,\rme^{-S[U]},
  \qquad
  \rmD[U]=\prod_{x,\mu}\rmd U(x,\mu),
  \enum
}
where ${\cal Z}$ is a normalization factor such that
$\langle 1\rangle=1$ and $\rmd U(x,\mu)$ the normalized invariant
measure on SU(3). To keep the
discussion as transparent as possible,
our attention will be restricted, in the following, to 
the Polyakov loop correlation function $\langle P(x)^{\ast}P(y)\rangle$
and to the expectation value of plane rectangular Wilson loops 
at $x_2=x_3=0$, with corners $(0,0),(t,0),(t,r),(0,r)$
in the $(x_0,x_1)$ plane.

\section 3. Factorization of the functional integral

In this section we rewrite the Wilson loop expectation values in a 
partly factorized
form that is closely related to the transfer matrix representation. 
Tensor products of pairs of link variables, 
the {\it two-link operators}, 
play an important r\^ole in this trans\-formation, 
and we thus introduce these first.
The factorization
reflects the local structure of the theory and 
will lead us (in sect.~4) to the multilevel simulation algorithm 
alluded to above.

\subsection 3.1 Two-link operators

The expectation value
of the rectangular loop defined at the end of the previous section
may be interpreted as a transition matrix element
between states that describe a pair of
static colour charges separated by a distance $r$ [\ref{Wilson}]. 
The charges are created at time $x_0=0$ and annihilated at $x_0=t$  
through the line operator 
\equation{
   \lineop(x_0)_{\alpha\beta}=
   \{U(x,1)\ldots U(x+(r-a)\hat{1},1)\}_{\alpha\beta},
   \qquad
   x=(x_0,0,0,0),
   \enum
}
and its complex conjugate respectively. 
Between these times the charge propagation
is represented by the two-link operators
\equation{
  \twolink(x_0)_{\alpha\beta\gamma\delta}=
  U(x,0)\cc_{\alpha\beta}
  \kern1pt
  U(x+r\hat{1},0)_{\gamma\delta}
  \enum
}
(see fig.~1). If we group the indices in pairs,
$(\alpha,\gamma)$ and $(\beta,\delta)$, these operators assume
the form of complex $9\times9$
matrices that act on colour tensors in the ${\bf 3}^{\ast}\otimes{\bf 3}$ 
re\-pre\-sen\-tation of SU(3). 

The multiplication of the time-like link variables in the Wilson loop
corresponds to the multiplication law 
\equation{
   \left\{\twolink(x_0)\twolink(x_0+a)\right\}_{\alpha\beta\gamma\delta}=
   \twolink(x_0)_{\alpha\lambda\gamma\epsilon}
   \twolink(x_0+a)_{\lambda\beta\epsilon\delta}
   \enum
}
for the two-link operators.
Using this rule the factorized expression
\equation{
  W({\cal C})=
  \lineop(0)_{\alpha\gamma}
  \left\{\twolink(0)\twolink(a)\ldots\twolink(t-a)\right\}
  _{\alpha\beta\gamma\delta}\lineop(t)\cc_{\beta\delta}
  \enum
}
is obtained, while in the case of a pair of Polyakov loops 
at distance $r$
along the $x_1$ axis, there are no line operators and 
the corresponding formula,
\equation{
  P(x)^{\ast}P(x+r\hat{1})=
  \left\{\twolink(0)\twolink(a)\ldots\twolink(T-a)\right\}
  _{\alpha\alpha\gamma\gamma},
  \enum
}
assumes an even simpler form.

\topinsert
\vbox{
\vskip0.0cm
\epsfxsize=5.5cm\hskip3.5cm\epsfbox{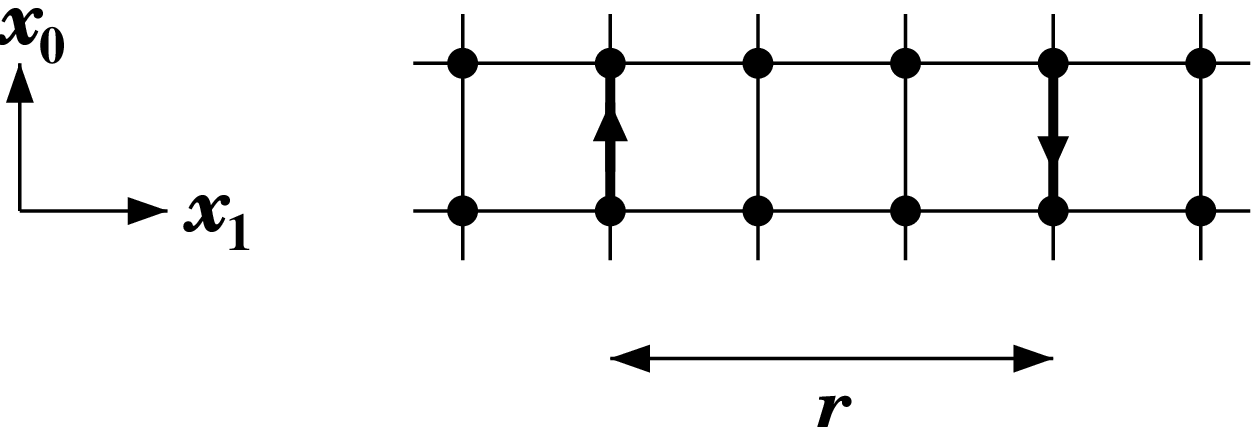
}
\vskip0.2cm
\figurecaption{%
The two-link operator $\stwolink(x_0)$
is equal to the tensor product of the
time-like link variables at $x=(x_0,0,0,0)$ and 
$x=(x_0,r,0,0)$.
}
\vskip0.0cm
}
\endinsert

\subsection 3.2 Sublattice expectation values

Averages over sublattices is the next topic that we need to discuss.
In the present context the relevant sublattices
are time-slices of variable thickness. Such a time-slice
consists of all lattice points 
with time coordinates in a given interval 
$[x_0,y_0]$ and its boundary are 
the equal-time hyperplanes at times $x_0$ and $y_0$. 

Lattice gauge theory on a time-slice can be studied independently of
the surrounding lattice if the spatial link variables
at the boundaries are held fixed. This decoupling property
is a consequence of the locality of the action (2.1) and more precisely
of the fact that the action involves plaquette loops only.
The link variables in the interior of the time-slice are then
the dynamical degrees of freedom 
that are to be integrated over when calculating
sublattice expectation values. We define the latter in the obvious
way, with the action reduced to those terms that depend
on the dynamical link variables. 
Sublattice expectation values are 
denoted by a square bracket $[\ldots]$ in order to distinguish 
them from the expectation values $\langle\ldots\rangle$ on the whole lattice.

Later we shall be dealing mostly with time-slice expectation values
of products of two-link operators. 
These are explicitly given by
\equation{
  \left[\twolink(x_0)\ldots\twolink(y_0-a)\right]
  ={1\over{\cal Z}_{\sub}}
  \int\rmD[U]_{\sub}
  \,\twolink(x_0)\ldots\twolink(y_0-a)\,
  \rme^{-S[U]_{\sub}},
  \enum
} 
where the index ``sub'' indicates that the sublattice expression is meant.
In particular, the sublattice partition function ${\cal Z}_{\sub}$
is determined by the requirement that $[1]=1$. 
It will be important in the following to keep in mind that 
sublattice expectation values are well-defined functions of the 
link variables at the boundary of the sublattice, but do not depend
on the gauge field elsewhere on the lattice.

\subsection 3.3 Hierarchical integration formulae

We now insert the product representation (3.4) in the 
expectation value of the Wilson loop and
reorganize the functional integral in a hierarchical way.
The manipulations are essentially the same as
in the case of the multihit method [\ref{MultiHit}], but they are
applied to the two-link operators and are carried to higher levels.

We first note that the expectation value can be
rewritten in the form
\equation{
  \left\langle W({\cal C})\right\rangle
  =\bigl\langle
  \lineop(0)_{\alpha\gamma}
  \left\{\left[\twolink(0)\ldots
  \twolink(t-a)\right]\right\}
  _{\alpha\beta\gamma\delta}\lineop(t)\cc_{\beta\delta}
  \bigr\rangle,
  \enum
}
where the square bracket stands for the expectation
value in the time-slice $[0,t]$. 
The Wilson loop expectation value may thus be computed
in two steps, first calculating the time-slice 
average of the product of the 
two-link operators, for the current configuration of the 
gauge field at the boundary of the time-slice, 
and then the full lattice average of the product of the 
time-slice expectation value with the line operators.

We can in fact derive various expressions of this type,
with many levels of nested averages. The key observation is
that the time-slice expectation values are compatible with each other
in the sense that they satisfy identities like
\equation{
  \left[\twolink(x_0)\twolink(x_0+a)\right]
  =\left[\left[\twolink(x_0)\right]
  \left[\twolink(x_0+a)\right]\right].
  \enum
}
The inner square brackets here refer to the time-slices $[x_0,x_0+a]$
and $[x_0+a,x_0+2a]$, respectively, 
while the time interval for the outer bracket
is $[x_0,x_0+2a]$.
A formula for the Wilson loop expectation value 
involving three levels of averaging, for example, is given by
(for $t/a$ even)
\equation{
  \left\langle W({\cal C})\right\rangle
  =\bigl\langle
  \lineop(0)_{\alpha\gamma}
  \left\{\left[\left[\twolink(0)\right]\left[\twolink(a)\right]\right]\ldots
  \left[\left[\twolink(t-2a)\right]\left[\twolink(t-a)\right]\right]\right\}
  _{\alpha\beta\gamma\delta}\lineop(t)\cc_{\beta\delta}
  \bigr\rangle.
  \enum
}
In general we may choose an arbitrary hierarchy 
of time-slices of increasing thickness
that fit within one another and with the time extent of the loop.
The pattern does not even need to be regular, although 
this will often be the case in practice.

\subsection 3.4 Centre symmetry \& quark confinement

On any time-slice $[x_0,y_0]$, the sublattice theory 
has a global $\gz_3$ symmetry
that acts on the time-like link variables according to
\equation{
  U(x,0)\to U(x,0)z, \qquad
  z=\rme^{i2\pi k/3},
  \qquad
  k=0,1,2.  
  \enum
}
The symmetry situation is thus similar to the one
in finite-temperature lattice gauge theory.
In particular, for fixed time-slice thickness
and small values of the gauge coupling,
the $\gz_3$ symmetry is probably spontaneously broken.
The time-slices should otherwise be in the confinement phase
and we then expect that
\equation{
  \left\|\left[\twolink(x_0)\ldots\twolink(y_0-a)\right]\right\|
  \propto
  \rme^{-m_0r}
  \enum
}
at large distances $r$, where $\|\ldots\|$ denotes
the usual operator norm for $9\times9$ matrices acting on complex
vectors.
This is surely so in the strong
coupling regime, and the picture is also supported, at least qualitatively, 
by our present (limited) numerical experience.

Combining eq.~(3.11) with the hierarchical integration 
formulae derived above,
it follows that the Wilson loop expectation value satisfies an 
area law with string tension $\sigma\geq m_0/(y_0-x_0)$. 
The decay properties of the time-slice expectation value (3.6) are thus
directly linked to the issue of quark confinement.

\section 4. Multilevel simulation

In the context of numerical simulations,
the hierarchical integration formulae derived above do not
seem to be particularly useful, because the time-slice
averages cannot in general be calculated exactly or
so precisely that the errors are surely negligible.
However, as in the case of the multihit method [\ref{MultiHit}], 
the sublattice averages may be estimated stochastically 
without compromising the correctness of the simulation.
Our aim in this section is to work this out in some detail
and to explain why the resulting multilevel algorithm may be 
expected to be highly efficient for large loops.

\subsection 4.1 Update cycle

Let us consider, as a simple example,
the Polyakov loop correlation function on a lattice with an even
number of points in the time direction.
A hierarchical integration formula
that might be used in this case is given by
\equation{
  \bigl\langle P(x)^{\ast}P(x+r\hat{1})\bigr\rangle=
  \bigl\langle\left\{\left[\twolink(0)\twolink(a)\right]\ldots
  \left[\twolink(T-2a)\twolink(T-a)\right]\right\}
  _{\alpha\alpha\gamma\gamma}\bigr\rangle.
  \enum
}
The associated multilevel simulation algorithm then proceeds
along the following lines:

\vbox{\parindent=1.8em
\vskip1.5ex
\item{(1)}{Generate a sequence of gauge field configurations
using a mixture of heatbath and over-relaxation link updates as 
usual.}

\vskip0.5ex
\item{(2)}{For a subsequence of configurations, estimate 
$[\twolink(x_0)\twolink(x_0+a)]$ by updating the gauge field
in the interior of the time-slice $[x_0,x_0+2a]$ a number of times and 
by averaging $\twolink(x_0)\twolink(x_0+a)$ over these
configurations.}

\vskip0.5ex
\item{(3)}{Compute the average of the trace of 
the product in eq.~(4.1) using the stochastic estimates for 
$[\twolink(0)\twolink(a)],\ldots,[\twolink(T-2a)\twolink(T-a)]$ 
calculated in step (2).}
}

\vskip0.5ex
\noindent
In practice the simulation proceeds sequentially, and 
the second step may be integrated in the first
by performing $n_1$ updates of the whole lattice,
then $n_2$ updates of the time-slices, then
$n_1$ full updates, and so on. 
The trace of the product of the time-slice expectation values
$[\twolink(x_0)\twolink(x_0+a)]$
is calculated in each of these cycles,
and the Polyakov loop correlation function is finally obtained
by computing the average of these values over 
a significant number $n_0$ of cycles. 

It should be emphasized that this algorithm 
produces exact results, up to statistical errors of order $(n_0)^{-1/2}$,
for any choice of $n_1\geq1$ and $n_2\geq1$. 
In this respect the quality of the stochastic estimation 
of the time-slice averages is therefore irrelevant,
but as we shall see shortly the efficiency of the simulation
strongly depends on it.
A formal proof of the exactness of the algorithm 
is given in appendix A.

\subsection 4.2 Exponential error reduction

We now show that the two-level algorithm described above 
leads to an exponential reduction of the statistical errors
if the time-slices of thickness 
$2a$ are in the confinement phase,
where the expectation values $[\twolink(x_0)\twolink(x_0+a)]$
decay exponentially at large $r$.

The stochastic
estimates of the latter, which are obtained in each cycle
of the algorithm, 
are accurate to within a statistical error proportional to 
$(n_2)^{-1/2}$. We may, for example, fix $n_2$ so that 
the signal-to-noise ratio is approximately equal to
unity at the specified value of $r$,
which requires $n_2$ to be scaled according to
\equation{
  n_2\propto\rme^{2m_0r}
  \enum
}
[cf.~eq.~(3.11)].
The factors in the product (4.1) are then of order $\rme^{-m_0r}$,
so that the magnitude of the trace of the product 
calculated in each cycle is roughly proportional to $\rme^{-m_0rT/2a}$.
In particular, the statistical fluctuations are reduced to this level.

As a result we expect that the algorithm achieves 
an exponential error reduction, 
with exponent proportional to the area $A=rT$,
for a computational effort growing as suggested by eq.~(4.2).
It goes without saying that
this analysis ignores 
many details and can only give a first indication of what
the true behaviour of the algorithm is going to be.

\subsection 4.3 Higher-order schemes

It should be quite clear at this point that any given hierarchical 
integration formula, with possibly many levels
of nested time-slice expectation values, corresponds to a multilevel 
simulation algorithm. At each level the associated time-slice expectation 
values are estimated stochastically, and these estimates
are then used in the averages taken at the next higher level.
The algorithm thus follows a cycle during which
the thin time-slices are updated more often than the thicker ones.

Further details on the algorithm and how to program it can be found in
appendix B. Here we only note that at the lowest level, where the products
of the basic link variables are averaged, the multihit method 
[\ref{MultiHit}] (perhaps also in its analytic version [\ref{DeFRo}])
may be
used to achieve a further reduction of the statistical errors.

\section 5. Test of the method

\vskip-4.0ex

\subsection 5.1 Lattice size and choice of parameters

For this first test of the multilevel simulation algorithm,
we decided to calculate the Polyakov loop correlation function
at gauge coupling $\beta\equiv6/g_0^2=5.7$, where 
the Som\-mer reference scale $r_0$ is about $2.92a$
[\ref{SommerScaleA},\ref{SommerScaleB}].
This implies $a\simeq0.17\,\fm$, and 
a lattice of spatial size $L=12a$ is thus approximately $2\,\fm$ wide.

Some experimenting reveals that
the time-slices of thickness $2a$ 
appear to be in the confinement phase, 
for this gauge coupling and lattice size, and if $T\geq6a$.
The two-level simulation algorithm discussed in the 
previous section is hence expected to perform well.
As it turns out,
for the larger values of $r$ and $T$, 
an even greater algorithmic efficiency
is achieved with a further level of averaging. 
The corresponding hierarchical integration formula reads
\equation{
  \bigl\langle P(x)^{\ast}P(x+r\hat{1})\bigr\rangle=
  \noenum
  \next{2ex}
  \qquad
  \bigl\langle\left\{
  \left[\left[\twolink(0)\twolink(a)\right]
        \left[\twolink(2a)\twolink(3a)\right]
  \right]\ldots
  \left[\ldots
        \left[\twolink(T-2a)\twolink(T-a)\right]
  \right]\right\}
  _{\alpha\alpha\gamma\gamma}\bigr\rangle,
  \enum
}
where $T/a$ is assumed to be a multiple of $4$.

\topinsert
\newdimen\digitwidth
\setbox0=\hbox{\rm 0}
\digitwidth=\wd0
\catcode`@=\active
\def@{\kern\digitwidth}
\tablecaption{Results for the Polyakov loop correlation function 
at $\beta=5.7$, $L=12a$}
\vskip-0.5ex
$$\vbox{\settabs\+x&%
                  xxxxx&&xxxxx&%
                  xx&xxxxxxxxxxxx&%
                  xxxxxxx&%
                  xxxxx&&xxxxx&%
                  xx&xxxxxxxxxxxx&%
                  xxx&x\cr
\thicktablerule
\vskip1ex
                \+& \hfill $T/a$ \hfill
                 && \hfill $r/a$ \hfill
                 && \hfill $\bigl\langle P^{\ast}P\bigr\rangle$\hfill
                 && \hfill $T/a$ \hfill
                 && \hfill $r/a$ \hfill
                 && \hfill $\bigl\langle P^{\ast}P\bigr\rangle$\hfill
                 &&  \cr
\vskip1.0ex
\thintablerule
\vskip1.5ex
  \+& \hfill $12$ \hfill
  &&  \hfill $2$ \hfill 
  &&  \hfill $6.46(2)\times10^{-5}$ \hfill
  &&  \hfill $6$ \hfill
  &&  \hfill $6$  \hfill
  &&  \hfill $2.48(2)@\times10^{-4}$\hfill 
  &\cr
  \+& \hfill $12$ \hfill
  &&  \hfill $3$ \hfill 
  &&  \hfill $4.93(3)\times10^{-6}$ \hfill
  &&  \hfill $8$ \hfill
  &&  \hfill $6$  \hfill
  &&  \hfill $9.53(13)\times10^{-6}$ \hfill
  &\cr
  \+& \hfill $12$ \hfill
  &&  \hfill $4$ \hfill 
  &&  \hfill $5.42(5)\times10^{-7}$ \hfill
  &&  \hfill $12$ \hfill
  &&  \hfill $6$  \hfill
  &&  \hfill $1.90(4)@\times10^{-8}$ \hfill 
  &\cr
  \+& \hfill $12$ \hfill
  &&  \hfill $5$ \hfill 
  &&  \hfill $7.06(9)\times10^{-8}$ \hfill
  &&  \hfill $16$ \hfill
  &&  \hfill $6$  \hfill
  &&  \hfill $3.91(7)@\times10^{-11}$ \hfill 
  &\cr
  \+& \hfill $$ \hfill
  &&  \hfill $$ \hfill 
  &&  \hfill $$ \hfill 
  &&  \hfill $20$ \hfill
  &&  \hfill $6$  \hfill
  &&  \hfill $ 8.23(16)\times 10^{-14}$ \hfill
  &\cr
\vskip1ex
\thicktablerule
}$$
\endinsert

Multilevel algorithms have many parameters and it is not easy
to find their optimal values. The most important parameters are 
the numbers of time-slice updates that are to be performed at each level.
Using an optimization strategy discussed in appendix B, 
we found that a good efficiency
at distance $r=6a$ is achieved with 96 update sweeps at the lowest level
(time-slices of thickness $2a$) and 10 sweeps at the next-to-lowest
level (thickness $4a$).
In addition, the application of the multihit method
at the lowest level proved to be beneficial.
The ratio of heatbath
to over-relaxation link updates is not critical and was set to $1:5$.
At smaller distances $r$,
it is generally better to reduce the numbers of 
time-slice updates, which reflects the fact that it does not pay 
to determine the time-slice averages accurately.

\topinsert
\vbox{
\vskip0.0cm
\epsfxsize=8.6cm\hskip1.5cm\epsfbox{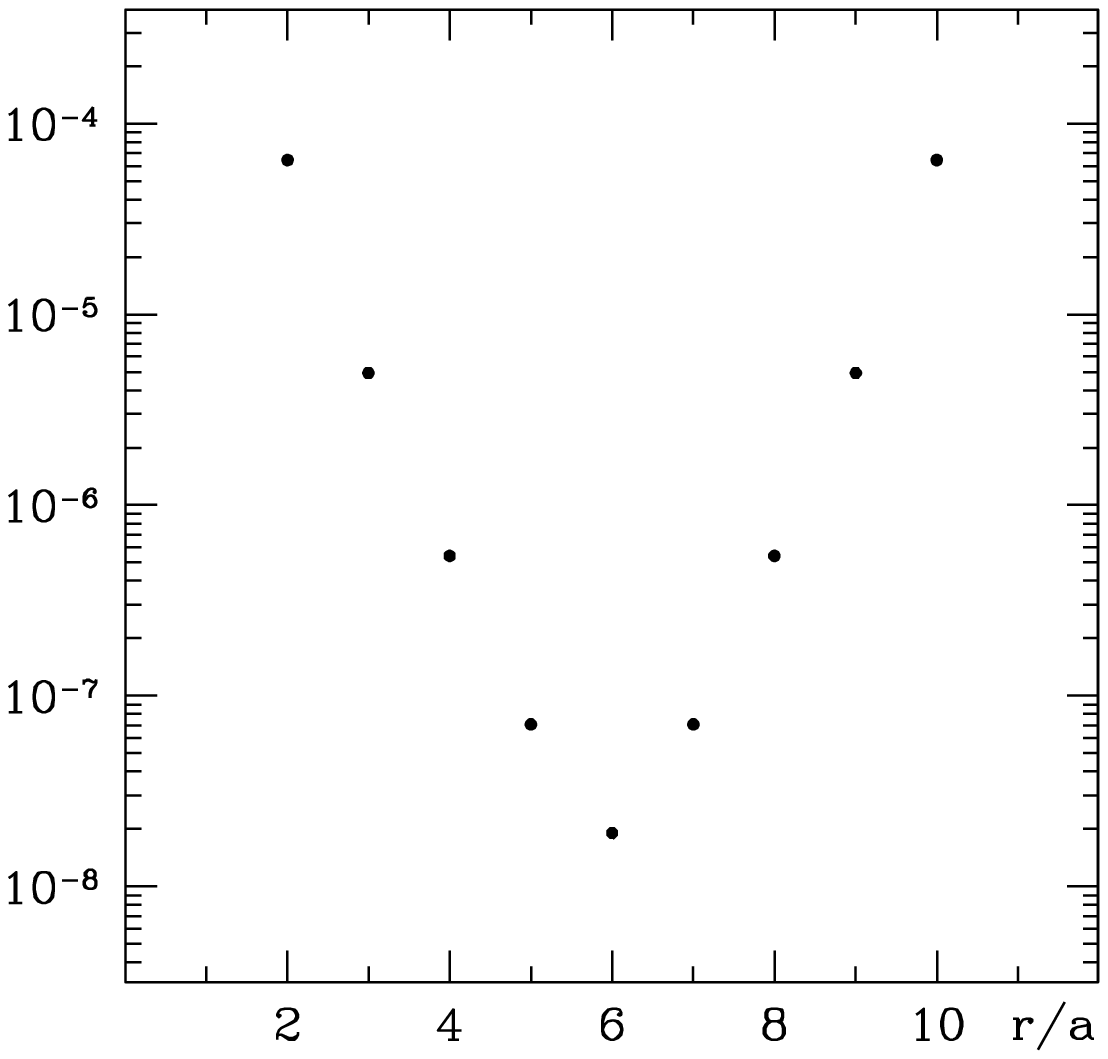}
\vskip0.0cm
\figurecaption{%
Polyakov loop correlation function at distance $r$ 
on a $12^4$ lattice at $\beta=5.7$. 
The loops on this lattice are about $2\,\fm$ long and enclose an area 
$A=rT$ ranging from $0.7$ to $2.1\,\fm^2$. 
Statistical errors are smaller than the dots.
}
\vskip0.0cm
}
\endinsert

\subsection 5.2 Simulation results

The results of our calculations (table~1 and figs.~2,3)
show rather strikingly that the Polyakov loop correlation function
is obtained for a large range of loop sizes and distances
with statistical errors that decrease exponentially.
With these algorithms the {\it relative}\/ 
errors can effectively be kept fixed
even if the correlation function is very rapidly decaying.

In total the simulations that we have done required the equivalent of 
about 1000 hours of processor time on a PC with 1.4 GHz Pentium 4 
processor.
With a program that makes use of the multihit method only,
and with any currently available super-computer,
it would be quite impossible
to reproduce the data listed in table~1.
At $r=6a$ and $T=12a$, for example,
the multilevel algorithm achieves an efficiency (in terms of the 
computer time required 
for a specified error on a given machine) that is better by 
a factor of $3\times10^5$ or so, and this factor
rises to astronomical values at the larger values of $T$.

\topinsert
\vbox{
\vskip0.0cm
\epsfxsize=10.0cm\hskip1.0cm\epsfbox{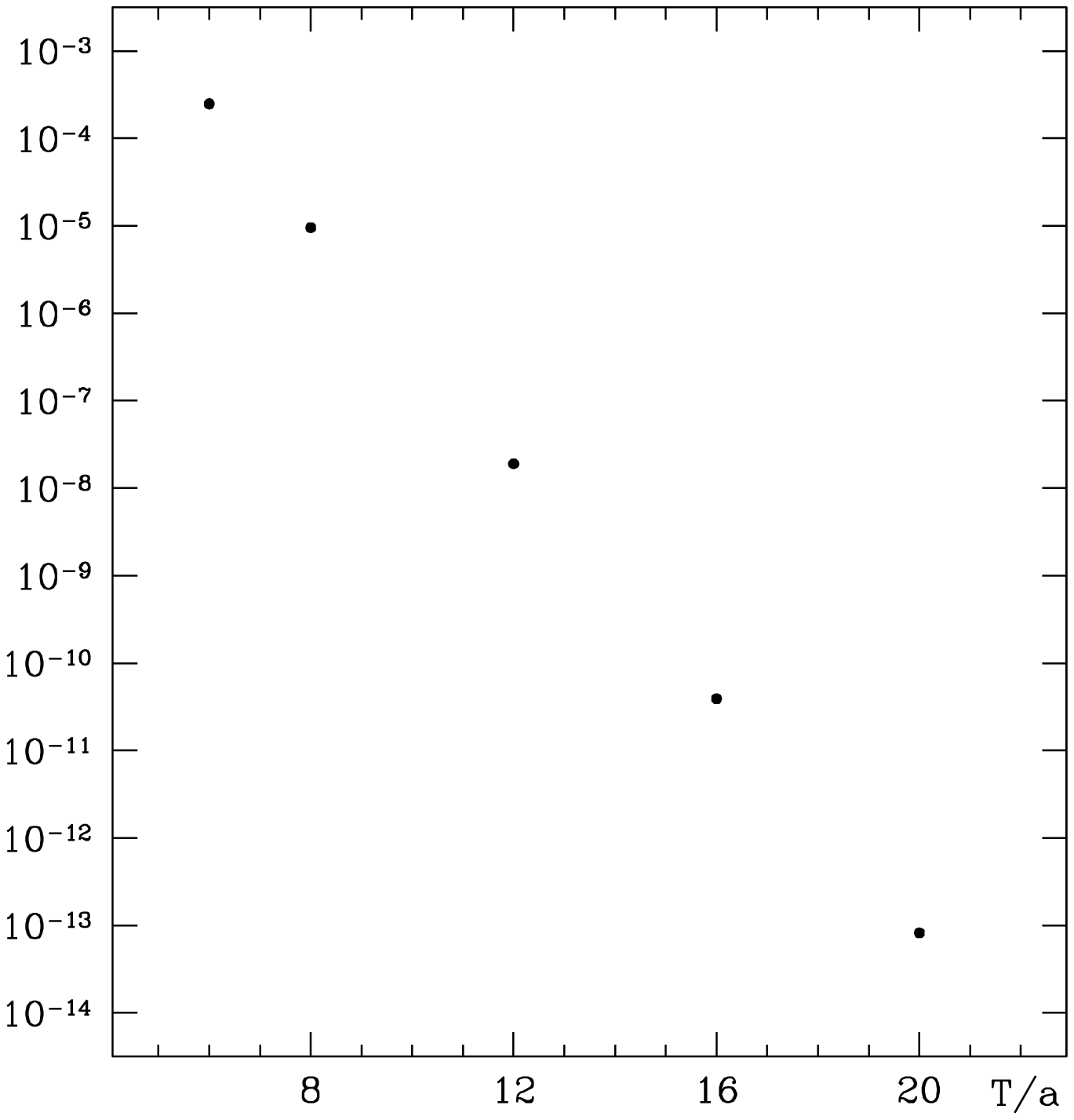}
\vskip0.0cm
\figurecaption{%
Polyakov loop correlation function at $\beta=5.7$ and 
distance $r=6a\simeq1.0\,\fm$, on a lattice of spatial size $L=12a$ 
and variable time extent $T$.
Statistical errors are not visible on this scale. On the 
largest lattice, $T=20a$, the loops are $3.4\,\fm$ long.
}
\vskip0.0cm
}
\endinsert

Our tests have also shown that the multilevel algorithm is
well behaved when the time extent of the lattice increases. 
For fixed $r$ and a given number
of ``mea\-sure\-ments'' of the loop correlation function, the relative
errors appear to be growing roughly linearly with $T$. 
In other words, the required computer time scales appro\-xi\-mately
like $T^3$ if the relative errors are to be kept fixed. 
Eventually this becomes a big factor, but the improvement
with respect to the exponential decrease of the signal-to-noise ratio
that is characteristic of the simulation algorithms used to date
is obvious.

\section 6. Conclusions

At the gauge coupling that we have considered,
the multilevel algorithm proposed in this paper performs exceedingly
well, and there is little doubt that a similarly impressive
error reduction will be achieved also at other couplings.
Using this technique, it is possible to compute loop 
correlation functions
and Wilson loop expectation values in a range of loop sizes 
and distances that had remained inaccessible so far. 
In particular, one of the physics issues that might be reconsidered 
at this point is the question of whether 
long colour flux tubes 
are described by an effective bosonic
string theory [\ref{StringA},\ref{StringB}] or 
perhaps a different kind of string theory 
(or none at all).

In some cases it may be useful to combine
the multilevel algorithm 
with link-blocking techniques [\ref{LBTeper},\ref{LBAlbanese}]
and the variational method 
[\ref{VMMichael},\ref{VMLuscher}]. The calculation 
of the energy spectrum of excited colour flux tubes
is an example for this. String breaking, on the other hand,
is perhaps better approached by considering correlation 
functions of long Polyakov loops, as we did in this paper,
since their exponential decay at large time extents $T$ 
is guaranteed to yield the true ground-state energy in the chosen charge
sector (i.e.~this is an unbiased method).

We finally mention that
the basic ideas underlying the algorithm 
(sublattice averages and hierarchical averaging) are 
likely to be more widely applicable.
An important case to consider are the
correlation functions of two or more local operators,
which are another instance where 
the signal-to-noise ratio is exponentially decreasing
if the established simulation techniques are employed.

\vskip1ex
PW sincerely thanks the Center for Computational Physics, Tsukuba,
for hospitality during the time that this work was completed.
All computations reported in this paper have been done at 
the computer centres of DESY at Hamburg and
Zeuthen. We would like to thank the staff of these centres
for their support.

\vskip-2ex plus 1ex

\appendix A

In this appendix we establish the correctness of 
the two-level simulation algorithm defined in sect.~4. 
The principal question is whether
the replacement of the two-link operators by their averages
over certain subsets of configurations is permissible.

\subsection A.1 Abstract model

To be able to bring out the essence of the argument more clearly,
an abstract system will be considered with
a finite number of states $s$. Each state is characterized by
a vector $(s_0,s_1,\ldots,s_n)$
of discrete variables, whose joint probability distribution 
is of the factorized form
\equation{
  p(s)=p_0(s_0)\prod_{k=1}^np_k(s_0,s_k),
  \enum
  \next{2ex}
  \sum_{s_0}p_0(s_0)=\sum_{s_k}p_k(s_0,s_k)=1.
  \enum
}
We are then interested in calculating expectation values
of factorized observables
\equation{
   {\cal O}(s)={\cal O}_0(s_0)\prod_{k=1}^n{\cal O}_k(s_0,s_k).
   \enum
}
Evidently this model 
fits the case of interest if we identify
$s_0$ with the space-like link variables at all even times,
$s_1$ with the link variables in the interior of the time-slice $[0,2a]$,
$s_2$ with those in $[2a,4a]$, and so on.

\subsection A.2 Two-level simulation

We now suppose that the system is simulated 
by updating the variables $s_0,s_1,\ldots,s_n$ 
one by one in an arbitrary order.
The update algorithm is assumed to be such
that the transition probability for changing $s_k$ at fixed $s_0$
is independent of the current values of all other variables.
This implies, in particular, that the algorithm
simulates, and thus preserves, 
the conditional probability distribution $p_k(s_0,s_k)$.

If we perform the simulation in cycles, where, in each cycle, 
$n_1$ updates of all variables are followed by $n_2$ updates
of $s_1,\ldots,s_n$ only, it is clear that 
the subsequence of configurations that are obtained
during the $n_1$ full updates simulates the probability distribution $p(s)$.
The complete sequence of configurations does that too, but 
it is useful to look at the $n_2$ intermediate configurations in a different 
way. Since $s_0$ is fixed in this part of each cycle,
the algorithm effectively generates
$n_2$ configurations of each $s_k$ separately, so that in total
there are $(n_2)^n$ configurations
of the vector $(s_1,\ldots,s_n)$.
In practice these are not stored in the memory of the computer at any time,
but for the theoretical discussion we may assume this 
to be the case.

\subsection A.3 Factorization lemma

As we just remarked,
the simulation generates
$(n_2)^n$ configurations of the vector $(s_1,\ldots,s_n)$
in each cycle, 
and we now focus on the sequence of these sets of states.

\proclaim Lemma A.1.
The vectors $(s_1,\ldots,s_n)$ in the sets with
the same value of $s_0$ occur with conditional probability
$p_1(s_0,s_1)\ldots p_n(s_0,s_n)$.

\noindent
To prove this, we first note that
the states $s$ 
at the end of the $n_1$ full updates in each cycle
are distributed with probability $p(s)$.
In particular, the states in this sequence with the same value of 
$s_0$ are distributed with probability
$p_1(s_0,s_1)\ldots p_n(s_0,s_n)$.

When the algorithm arrives at any of these points, it continues
to generate the next set of $(n_2)^n$ vectors 
with a transition probability that preserves the conditional
probabilities of the variables $s_k$.
Since the initial values of these variables 
are already properly distributed, and since 
they are statistically independent, the 
sequence of the sets of $(n_2)^n$
additional states with the specified value of $s_0$ 
must be distributed in the same way.

\subsection A.4 Expectation values

The expectation value of the factorized observable (A.3) 
can be written in the form
\equation{
  \langle{\cal O}\rangle=
  \sum_{s_0}p_0(s_0){\cal O}_0(s_0)\prod_{k=1}^n[{\cal O}_k](s_0),
  \enum
  \next{2ex}
  [{\cal O}_k](s_0)=
  \sum_{s_k}p_k(s_0,s_k){\cal O}_k(s_0,s_k).
  \enum
}
From lemma A.1 it now follows that the product
\equation{
  \prod_{k=1}^n[{\cal O}_k](s_0)
  \enum
}
is equal to the average over all sets of $(n_2)^n$ configurations
with the specified value of $s_0$ that occur in the course of 
the simulation. Moreover the average over the $(n_2)^n$
configurations
in any one of these sets is trivially given by the product of the 
averages of the factors ${\cal O}_k(s_0,s_k)$.
In particular, there is no
need to store any configurations other than the current one,
since the averages of the factors can be computed sequentially.

We have thus shown that the expectation value 
$\langle{\cal O}\rangle$ may be obtained by 
substituting stochastic estimates for the factors
$[{\cal O}_k](s_0)$ on the right-hand side of eq.~(A.4)
and by averaging the so calculated product over the sequence of values
of $s_0$.
For each $k$ the estimate of $[{\cal O}_k](s_0)$ 
is the average over the $n_2$ configurations
of $s_k$ that are generated in the second part of each update
cycle.

\appendix B

The implementation of the multilevel algorithm
requires some care in order to 
avoid excessive memory usage 
and arithmetic inefficiencies.
In the following paragraphs we discuss the structure of 
our program in outline and address some of the key points
that should be taken into account.

\subsection B.1 Basic update algorithm

We use the now standard ``hybrid over-relaxation'' simulation 
algorithm that combines heatbath with over-relaxation link updates
in an adjustable proportion (for a review, see ref.~[\ref{Kennedy}], 
for example). 
In both cases a link update involves
three Cabibbo--Marinari rotations [\ref{CabMar}] 
in the obvious SU(2) subgroups 
of SU(3). Depending on the driving force exerted by 
the surrounding link variables, the heatbath algorithm
of Creutz [\ref{Creutz}] or the one of 
Fabricius--Haan [\ref{HaanFab}] and Kennedy--Pendleton [\ref{KenPen}]
is applied. In this way a high efficiency is achieved in all situations.

It is well-known that the choice of the random number generator can 
introduce a systematic bias in numerical simulations. To be on the safe
side we use the {\tt ranlux} generator [\ref{RLXtheory},\ref{RLXprogram}] 
even though it consumes a significant fraction of the update time.

\subsection B.2 Program structure

The multilevel simulation algorithm cycles through several levels that 
correspond to time-slices of increasing thickness. 
At the lowest level the product
$\twolink(x_0)\ldots\twolink(y_0)$ 
is averaged over a set of configurations on the time-slice $[x_0,y_0]$. 
The result of this calculation
is then passed to the next level, where the contributions from two or more
lowest-level time-slices are multiplied and averaged. 
From here on the procedure
repeats itself until the top level is reached, at which point
the product of the nested averages of the two-link operators
is calculated and its trace is taken.

This algorithm thus has a tree-like structure, where at each level
the following parameters need to be specified:

\vskip1ex
\noindent
\quad$d_{\rm ts}$: thickness of the associated time-slice

\vskip0.5ex
\noindent
\quad$n_{\rm ms}$: number of ``measurements'' to be made for 
                   the time-slice average

\vskip0.5ex
\noindent
\quad$n_{\rm up}$: number of time-slice updates between ``measurements''

\vskip0.5ex
\noindent
\quad$n_{\rm hb}$, $n_{\rm or}$: numbers of heatbath and over-relaxation sweeps
per time-slice update

\vskip0.5ex
\noindent
\quad$p_{\rm ws}$: pointer to a memory area that may be used as workspace 
at this level

\vskip1ex
\noindent
In the program the full set of parameters is globally visible
so that the calculated time-slice expectation
values (which reside in the workspace of the associated level)
can easily be accessed when the algorithm 
has moved to the next higher level.

The level structure can be elegantly programmed by defining 
a recursive function that takes the level number
and the time-slice initial time $x_0$ as arguments and calculates
the corresponding (nested) time-slice average. Exactly what is to be 
done can be inferred from the globally visible parameters that describe
the hierarchy of the time-slices.
Internally the 
program calls itself until the lowest level is reached, and
the tree-structure of the algorithm is thus mapped to the 
call sequence generated by the program.

\subsection B.3 Rounding errors

In the range of loop sizes and distances that we have considered,
the Polyakov loop correlation function 
decays over many orders of magnitude.
One might suspect~that significance losses become an issue
when the calculated values approach the machine precision.
This is not the case, however, because the correlation function 
is obtained by averaging a product of factors of about
equal magnitude (that are themselves averages of products of 
still larger factors, etc.).
In other words, the small numbers 
do not result from an enormous cancellation but by multiplication
of many factors. 

Some significance losses may still occur when 
the time-slice averages are computed.
This problem (if present) 
can be avoided by performing all operations involving
two-link operators, their products and averages 
in 64-bit floating-point arithmetic. 
On the other hand, there is no reason not to use single precision
data and arithmetic for the link variables and the 
basic update algorithm.

\subsection B.4 Memory requirements

Since the update cycles of the multilevel algorithm 
are time-consuming, it is 
economical to calculate the Polyakov loop correlation function
simultaneously 
at all points $x=(0,x_1,x_2,x_3)$ and displacement vectors
$r\hat{k}$, $k=1,2,3$. 
At each level the workspace must then be sufficiently large to contain 
this many two-link operators. One actually needs 
twice this space for the calculation of the averages at all $x$.

Two-link operators have 162 real components and thus occupy 1296 bytes
of storage if double precision arithmetic is used.
The total memory space required per level is hence
$7.6\,\hbox{KB}\times(L/a)^3$
times the number of distances $r$ at which the correlation function 
is to be calculated. This quickly adds up to a large number,
but it should be taken into account that 
there is not much to be gained by processing many values of $r$
at the same time, because
the optimal choice of the level parameters depends on $r$.
Note that the same memory area may be used for all time-slices 
at a given level since these are visited sequentially.

\subsection B.5 Operation count and timing

The tensor product (3.2) and 
the product (3.3) of two two-link operators 
require 486 and 5670 floating-point 
operations respectively. These numbers are large but not out 
of proportion, considering the fact that 
1926 operations are required for the calculation of 
the ``staples'' in the link update programs.
In our test runs, for example,
the total time spent to manipulate the two-link 
operators was comparable to the time needed to update 
the gauge field.

At the lowest level, multiplications of two-link operators should
be avoided by first calculating the products $U(x_0,0)\ldots U(y_0-a,0)$
and then the tensor products of these. If the multihit method is
used, the link variables $U(z_0,0)$ are averaged
over a number of heatbath link updates before they are inserted in the 
products.

The simulations reported in sect.~5 
have been performed on an $8$-node cluster with 550 MHz Pentium III processors 
and on a stand-alone PC with 1.4 GHz Pentium 4 processor. 
Using vector arithmetic (SSE instructions), the 
processor time required for a heatbath (over-relaxation) link update
on this PC is $3.4\,\mu$s ($2.0\,\mu$s).
The timing of the multilevel algorithm is more difficult,
but a rough estimate of the execution time per cycle
may be obtained by adding the link update 
times and multiplying this number by 2.
On the $16\times 12^3$ lattice, for example,
100 ``measurements'' of the loop correlation function at distance $r=6a$,
with cycle parameters as quoted below,
require about 50 hours of PC processor time.

\subsection B.6 Parameter tuning

It is our experience that the parameters of the multilevel algorithm
are best determined by fixing them at the lowest level, 
then at the next-to-lowest level, and so on. Since the multihit method
leads to a further reduction of the statistical errors, the first step
is to optimize this part of the algorithm.
The thickness $d_{\rm ts}$ of 
the time-slice at the lowest level 
must then be determined. As discussed in subsects.~3.4 and 4.2, 
$d_{\rm ts}$ should be as small as possible, but 
sufficiently large that the time-slice is in the confinement phase. 

The other parameters listed in subsect.~B.2 are fixed 
essentially by minimizing 
the average over all points $x$ and directions $k$ of 
the {\it absolute value}\/ $|P^{\ast}P|$ of the trace of 
the product of the time-slice expectation values
calculated at the lowest level, for a single thermalized gauge field
configuration.
More precisely the average $\langle|P^{\ast}P|\rangle$
should be balanced against the processor time $\tau$ required to compute it
so that $\tau\times\langle|P^{\ast}P|\rangle^2$ is minimized (which yields
the maximal error reduction for a given amount of computer time).

At $\beta=5.7$, $r=6a$, $L=12a$ and all $T\geq8a$,
the lowest-level parameter list that we obtained in this way is 
$(d_{\rm ts},n_{\rm ms},n_{\rm up},n_{\rm hb},n_{\rm or})
=(2a,96,1,1,5)$.
The optimum is rather flat and variations of $n_{\rm ms}$ by 10\%
or so make practically no difference.
At the next level the same optimization procedure suggests to take 
$(d_{\rm ts},n_{\rm ms},n_{\rm up},n_{\rm hb},n_{\rm or})
=(4a,10,16,1,5)$.
The additional error reduction that is achieved at this level
is very substantial, but having a third level with $d_{\rm ts}=8a$
seems to have no positive effect.

We finally note that the auto-correlations between successive
``measurements" of the loop correlation function
can be practically reduced to zero by updating the full lattice
a significant number of times before every ``measurement".
This adds an only small overhead to the total execution time, 
which is dominated by the time required for 
the computation of the time-slice averages.


\ninepoint

\beginbibliography


\bibitem{MultiHit}
G. Parisi, R. Petronzio, F. Rapuano,
Phys. Lett. B128 (1983) 418


\bibitem{LBTeper}
M. Teper, 
Phys. Lett. B183 (1987) 345

\bibitem{LBAlbanese}
M. Albanese et al. (APE collab.),
Phys. Lett. B192 (1987) 163


\bibitem{VMMichael}
N. A. Campbell, A. Huntley, C. Michael, 
Nucl. Phys. B306 (1988) 51

\bibitem{VMLuscher}
M. L\"uscher, U. Wolff,
Nucl. Phys. B 339 (1990) 222


\bibitem{SBWittig}
O. Philipsen, H. Wittig, 
Phys. Rev. Lett. 81 (1998) 4056
[E: {\it ibid.} 83 (1999) 2684];
Phys. Lett. B451 (1999) 146

\bibitem{SBSommer}
F. Knechtli, R. Sommer, 
Phys. Lett. B440 (1998) 345
[E: {\it ibid.} B454 (1999) 399];
Nucl. Phys. B590 (2000) 309

\bibitem{SBForcrand}
O. Philipsen,  Ph. de Forcrand,
Phys. Lett. B475 (2000) 280


\bibitem{Wilson}
K. G. Wilson, Phys. Rev. D10 (1974) 2445


\bibitem{DeFRo}
Ph. de Forcrand, C. Roiesnel,
Phys. Lett. B151 (1985) 77


\bibitem{SommerScaleA}
R. Sommer,
Nucl. Phys. B411 (1994) 839

\bibitem{SommerScaleB}
M. Guagnelli, R. Sommer, H. Wittig, 
Nucl. Phys. B535 (1998) 389


\bibitem{StringA}
M. L\"uscher, K. Symanzik, P. Weisz,
Nucl. Phys. B173 (1980) 365

\bibitem{StringB}
M. L\"uscher,
Nucl. Phys. B180 (1981) 317


\bibitem{RLXtheory}
M. L\"uscher,
Comput. Phys. Commun. 79 (1994) 100

\bibitem{RLXprogram}
F. James,
Comput. Phys. Commun. 79 (1994) 111 [E: {\it ibid.}\/ 97 (1996) 357]


\bibitem{Kennedy}
A. D. Kennedy,
Nucl. Phys. B (Proc. Suppl.) 30 (1993) 96


\bibitem{CabMar}
N. Cabibbo, E. Marinari,
Phys. Lett. B119 (1982) 387

\bibitem{Creutz}
M. Creutz,
Phys. Rev. D21 (1980) 2308

\bibitem{HaanFab}
K. Fabricius, O. Haan,
Phys. Lett. B143 (1984) 459

\bibitem{KenPen}
A. D. Kennedy, B. J. Pendleton,
Phys. Lett. B156 (1985) 393

\endbibliography

\bye